# Software Reuse in Cardiology Related Medical Database Using K-Means Clustering Technique

M. Bhanu Sridhar[1], Y. Srinivas[2], M. H. M. Krishna Prasad[3]

[1]Department of Computer Science and Engineering, Raghu Engineering College, Visakhapatnam, India; [2]Department of Information Technology, GITAM University, Visakhapatnam, India; [3]Department of Information Technology, Jawaharlal Nehru Technological University, Kakinada, India.
Email: sridharbhanu@gmail.com, ysrinivasit@rediffmail.com, krishnaprasad.mhm@gmail.com

## ABSTRACT

Software technology based on reuse is identified as a process of designing software for the reuse purpose. The software reuse is a process in which the existing software is used to build new software. A metric is a quantitative indicator of an attribute of an item/thing. Reusability is the likelihood for a segment of source code that can be used again to add new functionalities with slight or no modification. A lot of research has been projected using reusability in reducing code, domain, requirements, design etc., but very little work is reported using software reuse in medical domain. An attempt is made to bridge the gap in this direction, using the concepts of clustering and classifying the data based on the distance measures. In this paper cardiologic database is considered for study. The developed model will be useful for Doctors or Para-medics to find out the patient's level in the cardiologic disease, deduce the medicines required in seconds and propose them to the patient. In order to measure the reusability K-means clustering algorithm is used.

**Keywords:** Reuse; Cardiology; Software Metrics; Clustering, K-Means; Cardiac

## 1. Introduction

Software Reuse is currently one of the most active and creative research areas in Software Engineering. It offers a solution to reduce repeated work and improve efficiency and quality in software development and management. It makes use of the experience obtained in the past development process. In the proposed article we have considered the database of the heart patients from [1] to focus on the cardiologic situations. Reuse is vital in medical field because the previous information is very handy in deducing a patient's current health position and save the precious life [2].

Cardiology is a medical specialty dealing with human heart disorders. This field includes diagnosis and treatment of disorders like heart defects, heart failure and other heart diseases. According to World Health Organization, India has the highest number of coronary heart disease deaths in the world [3]. This can be deduced not only due to lack of resources but also due to concentration of resources at places like cities and towns. By usage of Internet and cardiology database component reuse, the Para-medics, can deduce the medicines or methods to be used for the patients at remote places to temporarily put them out of danger. From the reuse of available data, the required medicines may also be deduced and proposed to the patients.

In this article we propose a methodology using the clustering technique together with classification technique where the heart patients' data is clustered, depending on the health conditions, into three categories: normal, pro-cardiac and cardiac. We use the Euclidean distance measure to classify the patients' disease level conditions into the three specified categories. The paper is organized as follows: Section 1 of the paper deals with introduction; in Section 2, categories of the heart patients is presented; K-means algorithm is presented in Section 3; Section 4 deals with the methodologies together with experimental results and finally the conclusion is presented in Section 5.

Our future work, which is at a research stage now would be very useful in aiding to the ailing patients and become an important part in the general usage of the Doctors.

## 2. Categories of Heart Patients

The heart is a myogenic (cell-related) muscular organ with a circulatory system (including all vertebrates), that is responsible for pumping blood throughout the blood vessels by repeated, rhythmic contradictions [4]. Among the problems related to heart, the major problem is car-diac arrest, which is the cessation of normal blood circulation due to failure of the heart to contract effectively. It should be effectively realized that cardiac arrest is different from a heart attack where blood supply is interrupted to a part of the heart which may/may not lead to the patient's death.

The patients who approach a doctor can be classified into three categories taking into consideration results of different tests conducted with the existing symptoms. The properties taken into consideration are Atherosclerosis (due to Cholesterol), Myocardial Infarction (heart attack), different medical signs like blood cell count and skin rashness, various symptoms like head ache and body pain, and other facts like Diabetes, Triglyceride, Migraine and so on [5].

Armed with all this information, the concerned patient is placed in one of the quoted three categories below.

### 2.1. Normal

A patient can be declared "normal" when no signs or symptoms of a cardiovascular/coronary disease are found within the results of various tests conducted. The general factors considered are the blood pressure (BP), sugar level in blood, results of Electrocardiography (ECG), Cholesterol level, Triglyceride, and other sensations. A normal patient should have the BP within control (<120/80 mmHg), Blood sugar level on waking up with an empty stomach between 80 to 120 mg/dl [6], normal output from the cardiac stress test conducted with the ECG [7], and no other notable problems. It should also be noted that a now normal patient might suffer from a heart stroke soon or later since he had inherited the problem, of which the reports wouldn't mention.

### 2.2. Pro-Cardiac

Pro-cardiac category keeps the account of those patients who are suspected to have some signs and/or symptoms of heart-problems. These can be observed from the BP tests slightly exceeding the normal levels, sugar levels in blood also rising, ECG suspecting (though not deducing) problems in future and some signs and symptoms like light chest pain, high cholesterol, severe head aches often turning up etc. do surface.

A pro-cardiac becomes a suspect of cardiac problems in near future and is advised by the Doctor not only to take a bit of medicines but also to consider doing regular exercises like light running, and other methods to bring his yet-controllable level to the normal state.

### 2.3. Cardiac

As it might be suspected, a cardiac is surely in the problematic range: prone to abnormal BP conditions, having severe pain the chest region, burning sensations, sweating, pain along the left arm and finally having already had a light heart attack. A cardiac must be immediately taken into consideration for regular treatment with constant observation of all concerned positions in and around the heart and those that affect the heart. A cardiac is also advised to taken high-power medicines and conduct long walks every day so as to keep the blood-pumping in the heart at a normal position.

After mentioning and discussing all the classification parts, it should also be noted that effective medical data of the patient should be readily available for the Doctors which also should be frequently updated. This data forms the backbone of the patient's classification level, severity level and the chance of saving his/her life. An attempt is made in this paper, by bringing into picture the reuse of data, to correctly judge the patient's position.

## 3. K-Means Clustering Algorithm

Clustering in data mining is the process of grouping a set of objects into classes of similar objects [8]. Many clustering algorithms are discussed in the literature and the most important of these are partitioning and hierarchical algorithms. K-means remains one of the most popular clustering algorithms used in practice [9]. The main reasons are it is simple to implement, fairly efficient, results are easy to interpret and it can work under a variety of conditions. The steps to be followed for effective clustering using K-means algorithm are:

**Step 1.** Begin with a decision on the value of $K$ = number of segments

**Step 2.** Put any initial partition that classifies the data into K segments. We can arrange the training samples randomly, or systematically as follows:

1) Take the first $K$ training samples as a single-element Segment.

2) Assign each of the remaining ($N$-$K$) training samples to the segment with the nearest centroid. Let there be exactly $K$ segments ($C1$, $C2$—$CK$) and $n$ patterns to be classified such that, each pattern is classified into exactly one segment. After each assignment, re-compute the centroid of the gaining segment.

**Step 3.** Take each sample in sequence and compute its distance from the centroid of each of the segments. If the sample is not currently in the cluster with the closest centroid switch this sample to that segment and update the centroid of the segment gaining the new sample and cluster losing the sample.

**Step 4.** Repeat step 3 until convergence is achieved, that is until a pass through the training sample causes no new assignments. After determining the final value of the $K$ (number of regions) we obtain the estimates the parameters $\mu_i$, $\sigma_i$ and $\alpha_i$ for the $i^{th}$ region using the segmented regions.

## 4. Methodology and Experimental Results

In this article a novel methodology for cardiac medical data reusability is proposed. A database from archives [11] is considered for carrying out our proposed work. In this method, we have first categorised the data into 3 groups namely, normal, pro-cardiac and cardiac. We have considered the scenario of Chintapalli, a remote tribal village in Andhra Pradesh, India, where no super-speciality services for treating cardiac patients are available. It is necessary in such conditions to supplement the patient with sufficient primary aid so that he can sustain for the minimum period of shifting. Depending upon the clinical reports of the patient's data, he is to be categorized into one of the levels presented in Sections 2.1-2.3. A dissimilarity matrix is constructed with the readings from the clinical observations and identifying the most leading factors that may be prone to the cardiac diseases as per the experts' references. The various readings considered are categorized into the above mentioned three groups and a database is formulated from the realistic data obtained from medical patients from the data referred in [10]. The predominant features considered in the database are: blood pressure (BP), heartbeat (HB), pulse rate (PR), ECG (normal/abnormal), pain in the left shoulder region, sweating, nausea/vomiting, over weight, chest pain and breathlessness.

For the testing purpose in this paper, we have used a database of ten patients with the above mentioned ten features; if the reading is present we have represented it by using a value 1 else 0 (binary). Following this procedure for the other inputs, a binary matrix [11] is obtained and this matrix is to be categorized; *K*-Means algorithm is utilized for the same. Now within the clusters, the homogenous data is obtained. To classify a patient, the dissimilarity matrix is again formulated and is classified by calculating the minimum distance between the posed query data and the retrieved data by using the clustering technique.

**Reuse Metrics**

The reuse components for partitioning the data are divided into 4 steps performed at each phase in preparation to the next phase. These steps are:

1) Developing a reuse plan or strategy after studying the problem and available solutions to the problem.

2) Identifying a solution structure for the problem following the reuse plan or strategy.

3) Reconfiguring the solution structure to improve the possibility of using predefined components available at the next phase.

4) Evaluating the system.

The major tasks under the first step are to understand the problem about the cardiac patients, build-up the knowledge for categorizing them into groups and develop a plan or strategy for their treatment. In the second step, apply the knowledge to develop a solution structure that is best suited for the problem following the reuse plan or strategy developed in the above phase. In the next step, reconfigure the solution in order to optimize the reuse both at both the current phase and next phase. Finally the computed components are to be classified using test features.

The data of 10 patients, from the archives [10] is converted into a binary matrix as above. The concepts in the clustering partition in reusable components [8] are utilized to construct a Java program that takes in the data from the **Table 1**. The program constructs the clusters by classifying the data using the Euclidean distance. After the K-Means clustering, the data is divided based on the binary clustering, into three groups. The patients with Ids (P4, P7, P3, P9, P10) belong to the first cluster, patients with Ids (P8, P2, P1) belong to the second cluster and patients with Ids (P5, P6, P10) belong to the third cluster.

The basic aim in this context is to assist the patients with minimum first aid for sustainability till he/she is shifted to the nearest multi-speciality clinic from the remote place Chintapalli considered here. In order to categorize the patients, it is necessary to identify the exactness of the category and thereby suggesting the minimum essential supportive drugs to maintain or better the current condition. It becomes clear by now that it is necessary to find the exactness of the disease if we are to achieve our goals.

To find the most exact solution in this concept, an auto-correlation model is used to find the exact correlation and categorization of the patients. The auto-correlation formula used here is given by

$$r_k = \frac{\sum_{t=k+1}^{n} (Y_t - \bar{Y})(Y_{t-k} - \bar{Y})}{\sum_{t=1}^{n} (Y_t - \bar{Y})^2}$$

where *t* is the patient with the first symptom, *K* + 1 is the patient with the second symptom and so on.

In this model, we try to correlate the data to each patient by considering the auto-correlation model and the results obtained are tabulated (**Figure 1**).

From the above considered data, it can be clearly seen that the patient with $R_6$ is having highest auto-correlation factor and is likely to have symptoms of a cardiac. The value obtained here is 0.9. The patient with Ids P5 and P6 *i.e.* R5 and R6 have the next immediate ranges and they are also likely to be cardiac-prone. The values obtained by using the above quoted autocorrelation formula are given under:

R1 = 0.3, R2 = 0.3, R3 = 0.1, R4 = 0.0023, R5 = 0.7, R6 = 0.9, R7 = 0.11, R8 = 0.3, R9 = 0.1, R10 = 0.72

Software Reuse in Cardiology Related Medical Database Using K-Means Clustering Technique 685

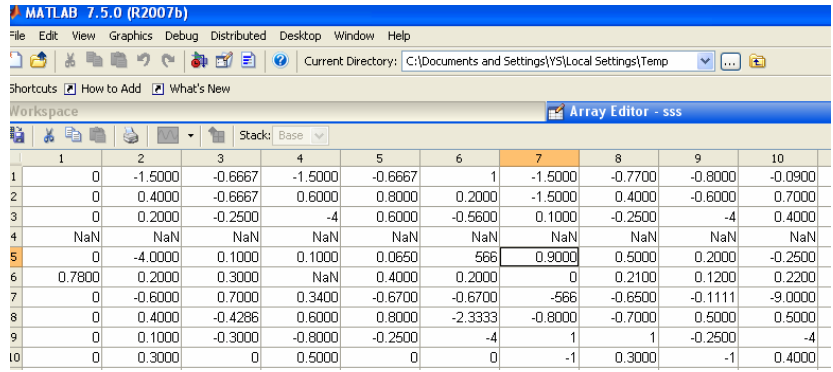

Figure 1. The results obtained from Autocorelation.

Table 1. The Symptoms (→) of the patients.

| Patient ID (↓) | BP | Heart beat (HB) | Pulse Rate (PR) | ECG | Left Shoulder pain | Sweating | Vomiting | Over Weight | Chest Pain | Breathlessness |
|---|---|---|---|---|---|---|---|---|---|---|
| P1 | 1 | 0 | 1 | 0 | 1 | 1 | 0 | 0 | 0 | 0 |
| P2 | 0 | 0 | 1 | 1 | 1 | 1 | 0 | 0 | 0 | 0 |
| P3 | 0 | 0 | 1 | 0 | 0 | 0 | 0 | 1 | 0 | 0 |
| P4 | 0 | 0 | 0 | 0 | 0 | 0 | 0 | 0 | 0 | 0 |
| P5 | 0 | 1 | 1 | 1 | 1 | 1 | 1 | 1 | 1 | 0 |
| P6 | 1 | 1 | 1 | 1 | 1 | 1 | 1 | 1 | 1 | 1 |
| P7 | 0 | 0 | 0 | 0 | 0 | 0 | 0 | 0 | 1 | 0 |
| P8 | 0 | 0 | 1 | 1 | 1 | 0 | 0 | 0 | 0 | 0 |
| P9 | 0 | 0 | 0 | 0 | 1 | 0 | 0 | 0 | 1 | 0 |
| P10 | 0 | 1 | 0 | 1 | 0 | 1 | 0 | 1 | 0 | 1 |

Here R6 is maximum, which specifies that the person is more likely to belong to the category cardiac; R1, R3, R4, R7, R8, R9 are at minimum risk and they belong to normal case and R2, R5 belong to the category pro-cardiac.

We have also tried to estimate the significance of each symptom for each patient over the other symptoms using auto-correlation and could identify the symptom that would be leading to cardiac problems.

We now input a new patient's data to check out the cluster where it belongs to; the Java program promptly supplies us the answer. The output of the Java program is given in **Figure 2**.

From the screenshot **Figure 2**, it can be easily identified that the given test data belongs to a particular cluster. Utilizing the classification given in Section 2, we obtain the concerned category.

## 5. Conclusion

In this paper a new methodology for software reuse in cardiac domain is presented. A database is considered or generated with 10 patients and is categorised into 3 categories depending upon the health conditions. The readings for these categories are obtained from the super speciality doctors, and are used for checking the reus-ability. The dissimilarity matrix is generated and the clustering is performed on the binary data. Classification is carried out on the test data by finding the minimum distance using Euclidean distance, and the reusability for partitioning is carried out as prescribed by Boris Delibasic *et al.* [8] are presented in Section 4.1.

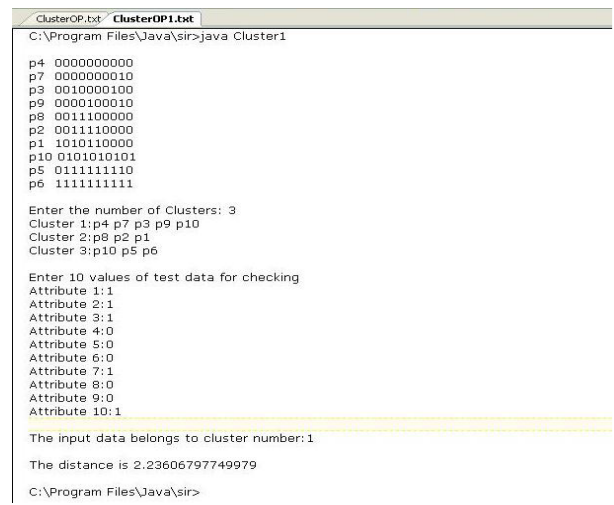

Figure 2. The results of classification.

The results obtained from the K-Means algorithm are given as inputs to the auto-correlation model to categorize the patients more accurately to be declared a cardiac. The model developed will be immensely useful for the Doctors to prescribe the medicines used for the previous patients of the respective cluster to the new patient immediately without spending time in checking conditions. It may be much more valuable for the Para-medics at remote places who can save the life of the patient.